\documentclass[a4paper,12pt]{article}
\usepackage[dvips]{graphicx}
\usepackage{epsfig}
\usepackage{graphicx}
\usepackage{color}
\usepackage{amssymb}
\usepackage{amsfonts}
\usepackage{amsmath}
\usepackage{xcolor}
\usepackage{hyperref}
\usepackage{color,soul}
 \soulregister\ref7
 \soulregister\cite7

\begin{document}
\title{Effect of coherent excitation in coherent electron cooler}
\author{S. Seletskiy\footnote{seletskiy@bnl.gov}, A. Fedotov, D. Kayran}
\date{\today}

\maketitle

\begin{abstract}
We consider the possibility of coherent excitation (CE) of cooled particles in the coherent electron coolers (CeC). We consider the current CeC scheme for the Electron Ion Collider (EIC) and derive the tolerances to a systematic error in longitudinal alignment of the electron and the proton bunches in the EIC cooler set by the CE effect.   
\end{abstract}

\section{Introduction}
In a conventional electron cooling \cite{BudkerInvent, BudkerDemo} the effect of coherent excitations \cite{Derbenev77} appears when the ion and the electron beams have a large enough difference in their average velocities. The friction force experienced by the cooled ions is a non-monotonic function of the relative ion-electron velocity. Therefore, in case of the friction force being shifted with respect to the zero (in the laboratory frame) ion velocity by an offset  larger than the velocity at which the first derivative of the friction force changes sign, a circular attractor appears in the ion bunch phase space. The effects that the presence of a circular attractor has on the dynamics of the ion bunch in the conventional electron coolers (both the non-relativistic coolers and the high energy ones) are discussed in \cite{circAttractLEReC}.

In this paper we will consider whether similar effect is expected in the EIC cooler based on the CeC technology \cite{CeCDerbenev, CeCLitvinenko, CeCRatner, CeCStupakov1, CeCStupakov2}. We will find that such an effect is possible and that it sets the tolerance to a longitudinal misalignment of the proton and the electron beams in the cooling section (CS) of the EIC cooler. Furthermore, we will show that even when the e-p alignment is perfect, a ``weak'' circular attractor might be present in the longitudinal phase space of the p-bunch under certain conditions. 

The parameters which we will use in our model are listed in Table \ref{param}.

\begin{table}[h!]
  \begin{center}
    \caption{Model parameters}
    \label{param}
    \begin{tabular}{lc}
    \hline
      $E_0$ (proton energy) [GeV] & 275 \\
      $V_0$ [eV] & 28\\
      $z_0$ [$\mu$m] & 6.7 \\
      $\sigma_0$ [$\mu$m] & 3 \\
      $R_{56}$ [mm] & 4  \\
      $N_p$ & $6.9 \cdot 10^{10}$ \\
      $\sigma_{\delta0}$   & $6.8 \cdot 10^{-4}$ \\
      $\sigma_{ps0}$ [cm] & 6 \\
      $\sigma_{es}$ [mm] & 4 \\
      \hline
    \end{tabular}
  \end{center}
\end{table}

\section{Equations of motion}

In this exercise we will use an approximation for the proton energy kick in the "kicker" section of the CeC found in \cite{wakeFormula}:

\begin{equation}\label{eq1}
    \Delta E(z)=-V_0 \sin \left( 2\pi \frac{z}{z_0} \right) \mathrm{exp} \left( - \frac{z^2}{\sigma_0^2} \right)
\end{equation}
where $V_0$, $z_0$ and $\sigma_0$ are the adjustable parameters (see Table \ref{param}) and $z$ is the proton longitudinal displacement in the kicker with respect to the proton's position in the ``modulator'' section of the cooler. We will assume that $z$ doesn't change through the kicker and that $z=R_{56} \delta$. Here, $\delta=\delta p/p_0$ is a relative momentum deviation of the proton and $R_{56}$ is the momentum compaction element of the modulator-kicker transfer matrix.

We will approximate the synchrotron motion of the protons by linear equations and we will use dimensionless variables $\delta$ and $\tau=\frac{\omega_s}{\eta}\frac{s}{\beta c}$, where $\omega_s$ is a synchrotron frequency, $\eta$ is a phase slip factor of the proton storage ring, $s$ is a proton's longitudinal position with respect to the center of the RF bucket, $\beta$ is a relativistic factor, and $c$ is the speed of light. Notice that $\sigma_{\delta0}=\frac{\omega_s}{\eta \beta c} \sigma_{ps0}$, where $\sigma_{\delta}$ and $\sigma_{ps}$ are respectively the rms momentum spread and the length of the ion bunch with the Gaussian distribution, and the index ``$0$'' signifies that here we chose to use the initial values of $\sigma_{\delta}$ and $\sigma_{ps}$. 

We introduce the invariant of the undisturbed motion $J=\delta^2+\tau^2$. For a bunch with the Gaussian distribution the density distribution function for $J$ is $f_J(J)=(1/\widetilde{J}) e^{-J/\widetilde{J}}$, where $\widetilde{J}=2\sigma_\delta^2$ is the average value of $J$. 

Finally, we will assume that, since the electron bunch length is much smaller than the length of  the p-bunch, and since the e-bunch is longitudinally placed at the center of the p-bunch, the protons interact with the electrons twice per synchrotron period when their synchrotron phase $\phi \approx \frac{\pi}{2}$ or $\frac{3\pi}{2}$.

Under these assumptions, the equations of motion of an individual proton are:

\begin{equation}\label{eq2}
\left\{
\begin{array}{ccl}
\tau' & = & \delta \\
\delta' & = & -\tau +\alpha \mathbb{C}(\phi) F(\delta)+ \sqrt{\alpha} \mathbb{C}(\phi) D  \\
F(\delta) & = & -\frac{V_0}{\beta^2 E_0} \sin \left( 2\pi \frac{R_{56} \delta-z_{off}}{z_0} \right) \mathrm{exp} \left( - \frac{(R_{56} \delta-z_{off})^2}{\sigma_0^2} \right)\\
D & = & \frac{V_0}{\beta^2 E_0} \sum \limits_i^{N_s} \left( e^{-a \varphi_i^2} \sin \varphi_i \right)
\end{array}
\right.
\end{equation}
Here $\tau' \equiv d\tau/d\phi$, $\delta' \equiv d\delta/d\phi$, $E_0$ is the proton beam energy, $z_{off}$ is a systematic longitudinal misalignment between the proton and electron bunches in the CeC kicker, and a Dirac comb function $\mathbb{C}$ is given by:

\begin{equation}\label{eq3}
\mathbb{C}(\phi)= \sum \limits_{n=0}^\infty \delta_D \left( \phi-\frac{\pi(2n+1)}{2} \right)
\end{equation}
where $\delta_D$ is a Dirac delta function.

The coefficient $\alpha$ is the number of times the proton lands on the e-bunch (having the rms length $\sigma_{es}$)when $\tau \approx 0$. The proton's slippage per revolution with a period $T_r$ is $\Delta s = \beta c \eta T_r \delta$. Therefore, $\alpha=\frac{2\sqrt{2\ln2}\sigma_{e}}{\Delta s}=\frac{\sqrt{2\ln2}\sigma_{es}\sigma_{\delta0}}{\pi Q_s \sigma_{ps0} |\delta|}$, where $Q_s$ is the synchrotron tune. The $\sqrt{\alpha}$ in front of the ``noise'' term $D$ signifies the random walk law.  

The diffusive term $D$ represents the sum of all the random kicks experienced by the particular proton from the wakes induced by $N_s$ protons in the slice. Here, we use notation $\varphi_i \equiv \frac{2\pi z_i}{z_0}$ and $a \equiv \left( \frac{z_0}{2\pi \sigma_0} \right)^2$. We chose the length of the slice to be $z_0$. Hence, $\varphi_i \in [-\pi, \pi]$ and, since by definition $\tau=s\cdot \sigma_{\delta0}/\sigma_{ps0}$, we get:

\begin{equation} \label{eqNs}
N_s=N_p z_0 \frac{\sigma_{\delta0}}{\sigma_{ps0}}\cdot f_\tau(0)
\end{equation}
where $N_p$ is the number of protons in the bunch, and $f_\tau(\tau)$ is the density distribution function for $\tau$. 

We assume that the proton bunch has a circularly symmetric distribution in $\tau,\delta$-coordinates. Let's notice that for our choice of variables $\tau=\sqrt{J}\cos(\phi)$. Let us also notice that for the given $\tau$ and $\phi$ (and for the corresponding  $J$) the protons' density in a $d\phi$ angle is $\frac{d\phi}{2\pi}f_J(J)$. Then the $\tau$-density distribution is represented by:

\begin{equation}\label{eqTau}
f_\tau(\tau)=\frac{1}{2\pi}\int \limits_{-\pi/2}^{\pi/2}\frac{2\tau}{\cos^2 \phi} f_J\left( \frac{\tau^2}{\cos^2 \phi} \right) d\phi
\end{equation}

For a p-bunch with the Gaussian distribution ($f_J(J)=(1/\widetilde{J}) e^{-J/\widetilde{J}}$) Eq. (\ref{eqTau}), as expected, gives $f_\tau(\tau)=\frac{1}{\sqrt{2\pi} \sigma_\delta} e^{-\frac{\tau^2}{2\sigma_\delta^2}}$, which results in $f_\tau(0)=\frac{1}{\sqrt{2\pi} \sigma_\delta}$ and $N_s = \frac{z_0 N_p \sigma_{\delta0}}{\sqrt{2\pi} \sigma_{ps0} \sigma_\delta}$.

Notice that the choice of $z_0$ for the slice length is reasonable for the case we are considering in this paper, since $z_0>\sigma_0$. For other cases $\sigma_0$ might be a better choice for the slice length because depending on the mechanism used to produce the kick (\ref{eq1}) it is possible to have $\sigma_0 \gg z_0$.

From (\ref{eq2}), noticing that averaging over $\phi$ zeros all terms $\propto D$, we get the ``average'' difference equation for $\delta^2$ over the synchrotron period $T_s$:

\begin{equation}\label{eq3a}
\frac{\Delta\delta^2}{\Delta t} = \frac{2}{T_s} \left(-\frac{\alpha_1V_0}{E_0}(\Phi(\delta,z_{off})+\Phi(\delta,-z_{off}))+\langle D^2 \rangle \frac{\alpha_1}{|\delta|} \right)
\end{equation}
where $\alpha_1 \equiv \alpha\cdot|\delta| = \frac{\sqrt{2\ln2}\sigma_{es}\sigma_{\delta0}}{\pi Q_s \sigma_{ps0}}$, and

\begin{equation}\label{eq3b}
\begin{aligned}
\langle D^2 \rangle &=  \left( \frac{V_0}{\beta^2 E_0} \right)^2 \frac{N_s}{2\pi} \int \limits_{-\pi}^\pi e^{-2a\varphi^2} \sin^2 \varphi d\phi \approx \left( \frac{V_0}{E_0} \right)^2 \frac{N_s}{2\pi} \int \limits_{-\infty}^\infty e^{-2a\varphi^2} \sin^2 \varphi d\phi =\\
 & =\left( \frac{V_0}{E_0} \right)^2 N_s \frac{\sqrt{\pi} \sigma_0}{2\sqrt{2} z_0} \left( 1 - e^{\frac{8\pi^2\sigma_0^2}{z_0^2}} \right) \approx \left( \frac{V_0}{E_0} \right)^2 N_s \frac{\sqrt{\pi} \sigma_0}{2\sqrt{2} z_0}
\end{aligned}
\end{equation}
A dimensionless function $\Phi(\delta, z_{off})=\sin \left( 2\pi \frac{R_{56} \delta-z_{off}}{z_0} \right) e^{\left( - \frac{(R_{56} \delta-z_{off})^2}{\sigma_0^2} \right)}$ was introduced in  Eq (\ref{eq3a}). Notice that since for a non-zero offset $z_{off}$ the energy kick becomes asymmetric with respect to $\delta=0$, one must consider the two semi-periods of synchrotron oscillation individually. This results in $(\Phi(\delta,z_{off})+\Phi(\delta,-z_{off}))$ expression in Eq (\ref{eq3a}).

Since in our model the p-e interaction is happening only at $\phi \approx \frac{\pi}{2}$, $\frac{3\pi}{2}$, at the moment of the interaction $J=\delta^2$. Therefore, combining equations (\ref{eqNs})-(\ref{eq3b}), and after averaging over the whole ensemble of the protons, we get:

\begin{equation}\label{eq3c}
\begin{array}{ccl}
\frac{1}{\widetilde{J}}\frac{d\widetilde{J}}{dt}&=&-\lambda_C+\lambda_D\\

\lambda_C&=&\frac{2\sqrt{2\ln2} V_0 \sigma_{es} \sigma_{\delta0}}{\pi T_r E_0 \sigma_{ps0}} \frac{1}{\widetilde{J}} 
\int \limits_0^\infty f_J (\Phi(\sqrt{J},z_{off})+\Phi(\sqrt{J},-z_{off})) dJ\\

\lambda_D&=&  \frac{\sqrt{\ln2} V_0^2 \sigma_{es} \sigma_0 \sigma_{\delta0}^2 N_p}{\sqrt{\pi} T_r  E_0^2 \sigma_{ps0}^2}
\cdot f_\tau(0) \cdot \frac{1}{\widetilde{J}} \int \limits_0^\infty \frac{f_J}{\sqrt{J}} dJ
\end{array}
\end{equation}
where $\lambda_C$ and $\lambda_D$ are the instantaneous cooling and heating rates respectively.

We found the equations defining the beam dynamics of the proton bunch with an arbitrary  distribution (as long as the distribution stays circularly symmetric in $\delta$, $\tau$ phase space).

Equation (\ref{eq3c}) gives the instantaneous cooling and heating rates for the bunch. It also defines the parameters of the equilibrium distribution given by $\lambda_C=\lambda_D$.

Equations (\ref{eq2}) are the proper equations of longitudinal motion for each proton. For practical simulations one must substitute $D$ in (\ref{eq2}) with the random kicks with the rms amplitude $\sqrt{\langle D^2 \rangle}$ given by (\ref{eq3b}) with $N_s$ and $f_\tau$ given by (\ref{eqNs}) and (\ref{eqTau}).

We finish this chapter by finding the cooling and heating rates for the p-bunch with the Gaussian distribution (which, of course, includes the case of the initial rates in the EIC cooler), and the zero longitudinal offset ($z_{off}=0$). For the Gaussian bunch the integrals in (\ref{eq3c}) (including $f_\tau$) can be taken analytically and we obtain:

\begin{equation}\label{eq3d}
\begin{array}{ccl}
\lambda_C&=&\frac{4\sqrt{\pi \ln2} V_0 \sigma_{es}}{T_r E_0 \sigma_{ps}} 
\frac{R_{56} \sigma_0^3}{z_0 (2R_{56}^2 \sigma_\delta^2+\sigma_0^2)^{3/2}} 
\exp \left( -\frac{2 (\pi R_{56} \sigma_0 \sigma_\delta)^2}{z_0^2 (2R_{56}^2 \sigma_\delta^2+\sigma_0^2)} \right)\\

\lambda_D&=&  \frac{\sqrt{\ln2} V_0^2 \sigma_{es} \sigma_0  N_p}{4\sqrt{\pi} T_r  E_0^2 \sigma_{ps}^2 \sigma_\delta^2}
\end{array}
\end{equation}

\section{Coherent excitations and development of circular attractor}

\subsection{Physics leading to formation of circular attractor}

For the sake of clarity, let us first ignore the diffusive term in (\ref{eq2}) and consider only the coherent ``not-noisy'' driving force $F$.

We notice that the longitudinal ``cooling force'' acting on the proton is a non-monotonic function of $\delta$. Hence, in analogy to the conventional electron cooling, we expect that the circular attractor in protons' longitudinal phase space can be formed if $z_{off}>z_1$, where $z_1$: $d\Delta E/dz |_{z_{1}}=0$. The ``critical offset'' $z_1$ is found from:

\begin{equation}\label{eq4}
 \frac{z_1}{\sigma_0^2}=\frac{\pi}{z_0}\cot \left( \frac{2\pi}{z_0} z_1 \right)
\end{equation}

For the parameters listed in Table \ref{param}, $z_1 \approx 1.33$ $\mu$m. 

\begin{figure}[!htb]
  \centering
  \includegraphics[width=0.7\columnwidth]{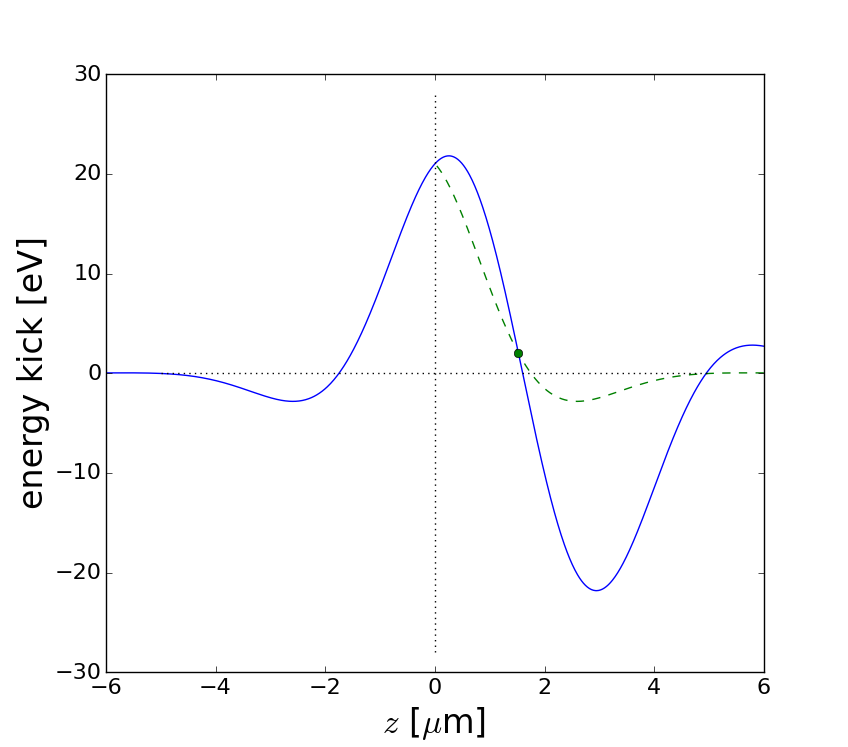}
  \caption{The energy kick experienced by a proton on a single pass through the CS (blue solid line). The green dashed line represents a reflection (around the vertical axis) of the kick function for negative $z$. The green dot shows a ``radius'' ($z_A$) of the attractor's projection on the z-space.}
   \label{energKick}
\end{figure}

Figure \ref{energKick} explains the physics underlying the attractor formation. For a nonzero offset the kick excites synchrotron oscillations for $R_{56} \delta  \in [0, z_{off}]$. The same proton passing the CS with negative $\delta$ will experience a damping kick. When $z_{off}>z_1$, there is a range of $\delta \in [-\delta_A, \delta_A]$ such that the exciting kick is larger than the damping kick. Hence, the average force acting on the protons with momentum deviations in this range is an exciting one. For protons with $\delta$ outside of this range the average force damps the synchrotron oscillations.  Therefore, instead of driving the amplitudes of proton oscillations to zero the cooling force with a sufficient enough offset makes all the protons to oscillate with the same non-zero amplitude. In other words, for $z_{off}>z_1$ the cooling force creates a circular attractor in the longitudinal phase space. The ``radius'' of the attractor's projection on $\delta$-space ($\delta_A=z_A/R_{56}$) is defined by:

\begin{equation}\label{eq5}
\begin{aligned}
\sin \left( 2\pi \frac{R_{56}\delta_A-z_{off}}{z_0} \right) \mathrm{exp} \left( - \frac{(R_{56}\delta_A-z_{off})^2}{\sigma_0^2} \right) & + \\
\sin \left( 2\pi \frac{R_{56}\delta_A+z_{off}}{z_0} \right) \mathrm{exp} \left( - \frac{(R_{56}\delta_A+z_{off})^2}{\sigma_0^2} \right) & = 0
\end{aligned}
\end{equation}

\begin{figure}[!htb]
  \centering
  \includegraphics[width=0.7\columnwidth]{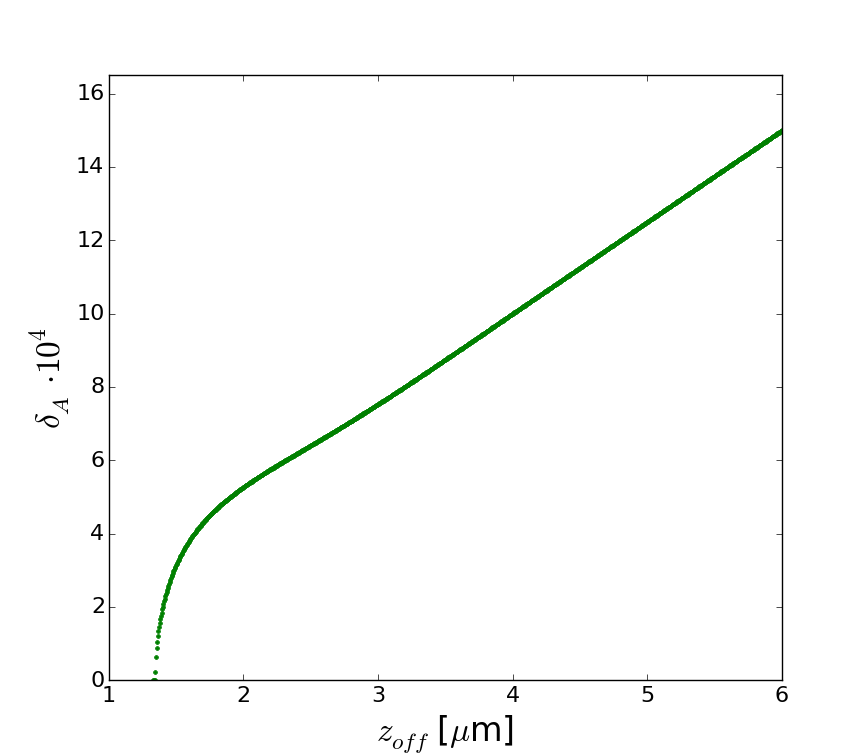}
  \caption{Numerical solution of Eq. (\ref{eq5}) for EIC cooler parameters.}
   \label{deltaZTrend}
\end{figure}

Since, the size of a circular attractor defines the minimum phase space volume to which the proton bunch can be cooled, it is useful to know a dependence of $\delta_A$ on $z_{off}$. Figure \ref{deltaZTrend} shows the solution of Eq. (\ref{eq5}) for the parameters listed in Table \ref{param}.

To simulate the individual proton dynamics (we are still ignoring the $D$-term) we integrate Eq. (\ref{eq2}) numerically with an explicit, exactly simplectic, third order method \cite{Ruth}. To make our simulations faster, and without a loss of generality, we increase the cooling force ($F(\delta)$) by a factor of $10^5$. 

\begin{figure}[!htb]
  \centering
  \includegraphics[width=0.9\columnwidth]{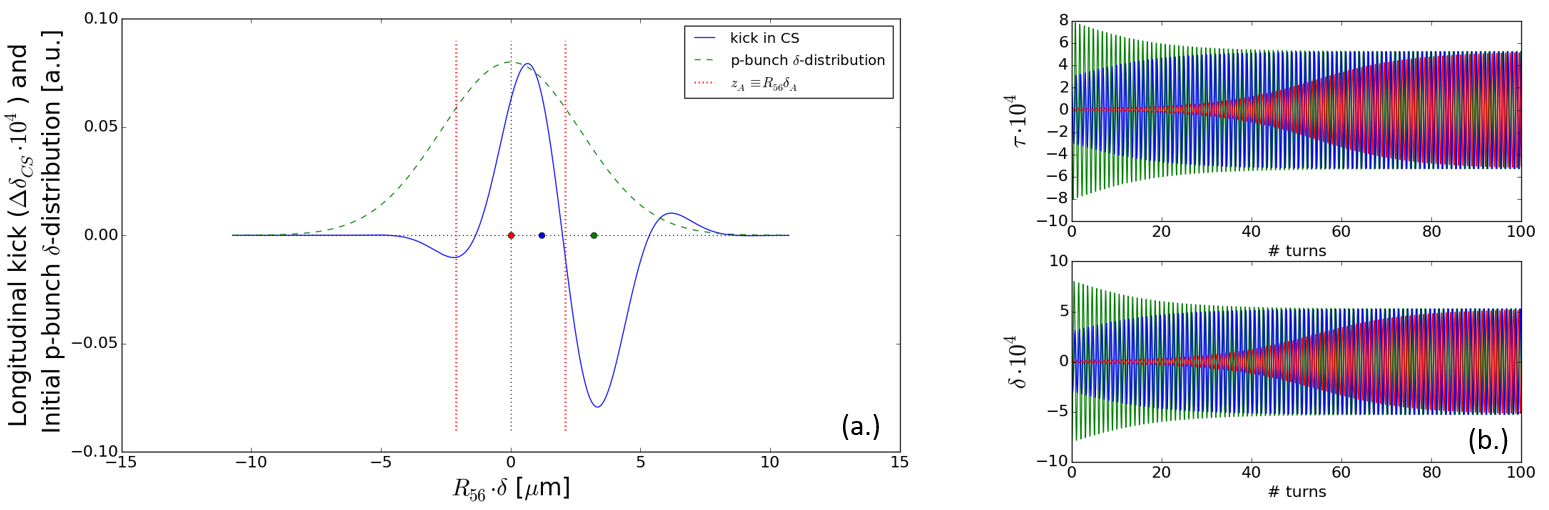}
  \caption{Plot (a) shows the longitudinal kick, the $\delta$-distribution of the p-bunch (with the rms $\sigma_\delta$ given in Table \ref{param}) and the ``size'' ($z_A$) of the attractor for $z_{off}=2$ $\mu$m. The red, the blue and the green dots in plot (a) show the initial $\delta$-amplitudes of the synchrotron oscillations of the three test protons. Plot (b) shows the evolution of synchrotron motion for the three test protons.}
   \label{offset}
\end{figure}

\begin{figure}[!htb]
  \centering
  \includegraphics[width=0.9\columnwidth]{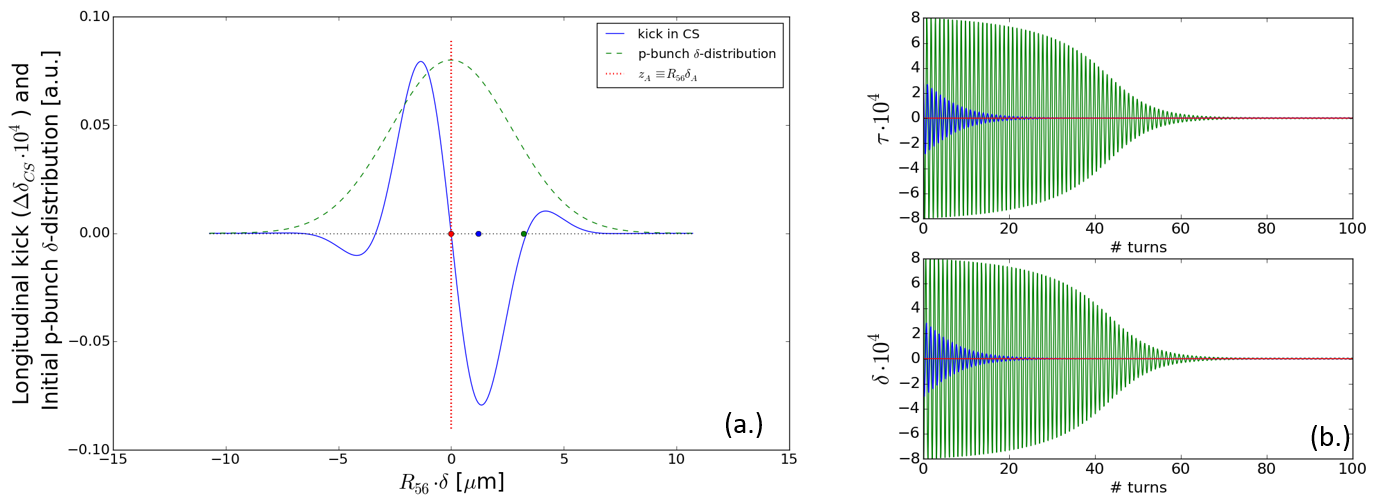}
  \caption{Plot (a) shows the longitudinal kick and the $\delta$-distribution of the p-bunch for $z_{off}=0$ $\mu$m. The red, the blue and the green dots in plot (a) show the initial $\delta$-amplitudes of the synchrotron oscillations of the three test protons. Plot (b) shows the evolution of synchrotron motion for the three test protons.}
   \label{noOffset}
\end{figure}

For example, Fig. \ref{offset} shows the evolution of the synchrotron oscillations of the three test protons in case of $z_{off}=2$ $\mu$m. For comparison, the evolution of the synchrotron motion of the same three protons for the case of $z_{off}=0$ $\mu$m is shown in Fig. \ref{noOffset}.

\begin{figure}[!htb]
  \centering
  \includegraphics[width=0.5\columnwidth]{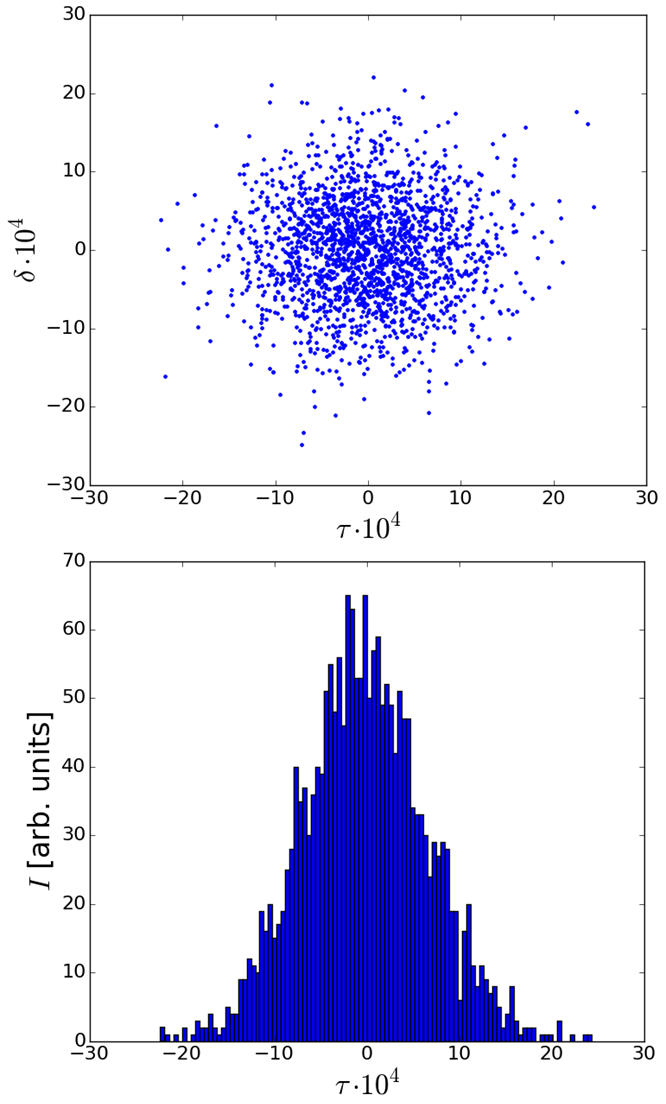}
  \caption{Initial proton bunch}
   \label{initialDistr}
\end{figure}

Next, we consider a bunch of (just 2000) protons with the Gaussian initial distribution in both $\delta$ and $\tau$ (Fig. \ref{initialDistr}). Then, we let it evolve for 100 synchrotron turns under the influence of the cooling force with $z_{off}=2$ $\mu$m. 

\begin{figure}[!htb]
  \centering
  \includegraphics[width=0.5\columnwidth]{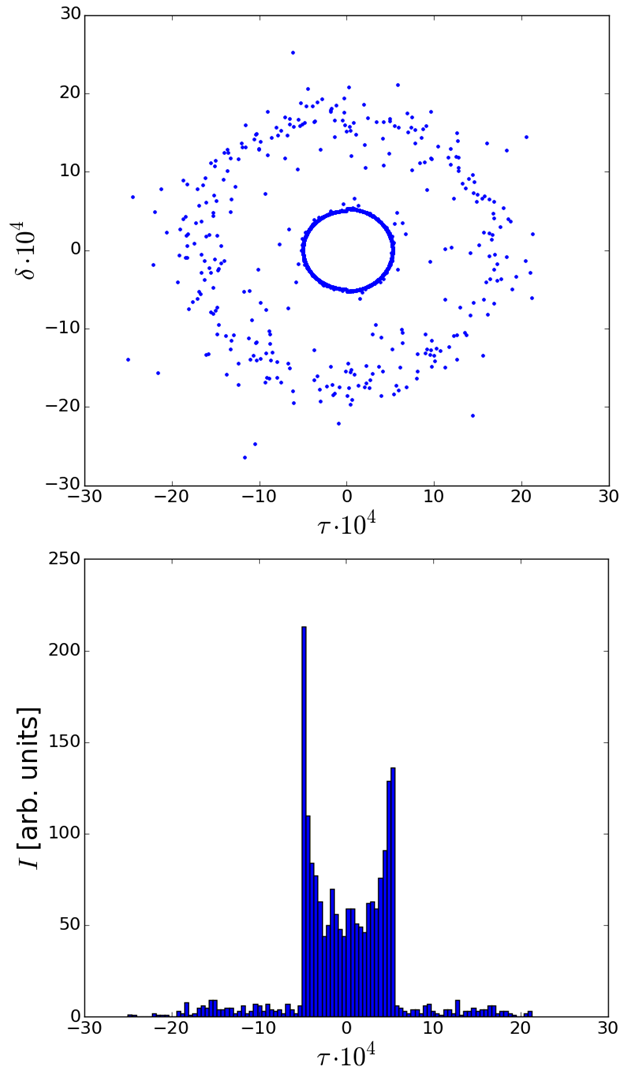}
  \caption{Phase space of the proton bunch after 100 synchrotron turns under the influence of the force with $z_{off}=2$ $\mu$m.}
   \label{strongAttractor}
\end{figure}

The result is shown in Fig. \ref{strongAttractor}. As expected, we see a ``strong'' attractor with $\delta_A$ defined by Eq. (\ref{eq5}). Yet, we notice that there is also another ``weak'' attractor. 

\begin{figure}[!htb]
  \centering
  \includegraphics[width=0.5\columnwidth]{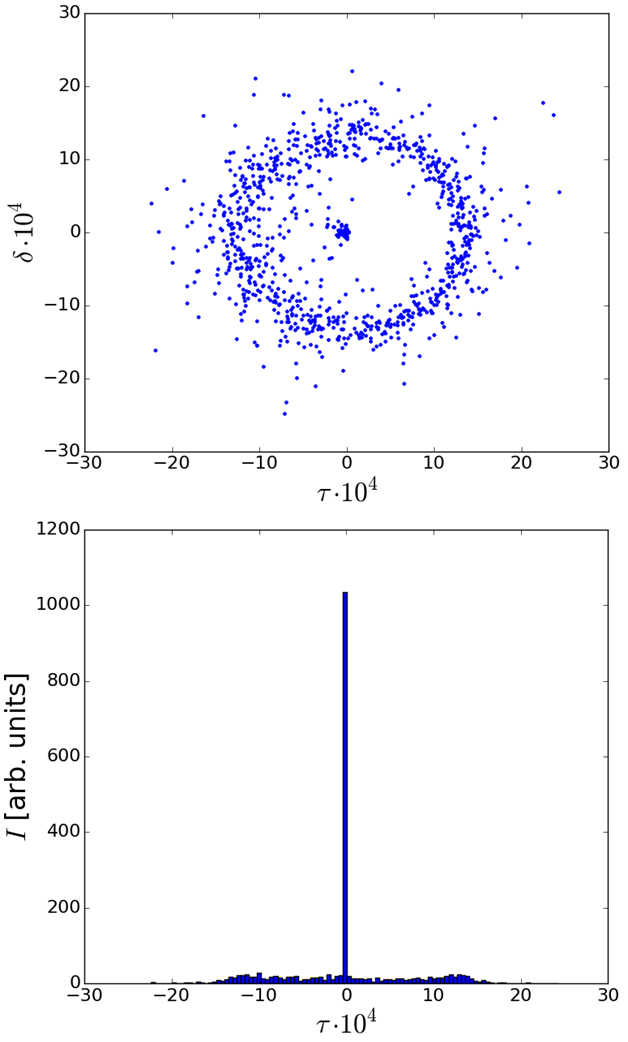}
  \caption{Phase space of the proton bunch after 100 synchrotron turns under the influence of the force with $z_{off}=0$ $\mu$m.}
   \label{weakAttractor}
\end{figure}

If we simulate the evolution of the p-bunch phase space in the presence of the cooling force without the offset we see that this weak attractor is still present, as Fig. \ref{weakAttractor} demonstrates.

The weak attractor exists because the ``cooling'' force has an actual ``heating'' portion and because in our model the protons interact with the e-bunch only when the longitudinal proton velocity is at its extremum (when $\phi=\frac{\pi}{2}$ or $\frac{3\pi}{2}$). Therefore, in our model the weak attractor must appear, even in the absence of an offset, at $z=z_0$ where the wake crosses zero the second time.

It is important to notice, that although the model wake (Eq. (\ref{eq1})) does not give a good numerical approximation of the actual wake in the kicker for large $z$, it does have the same qualitative behavior as the real wake function. Also, for the e-bunch being much shorter than the p-bunch, our assumption of p-e interaction happening only for  $\phi \approx\frac{\pi}{2}$ or $\frac{3\pi}{2}$ will be satisfied for a large portion of the proton bunch. Therefore, we conclude that the weak attractor can appear in the realistic EIC cooler.

Whether the IBS-driven diffusion comparable in its strength to the cooling force will ``shake'' the protons off the weak attractor remains to be explored. If the IBS won't help with suppressing the weak attractor then one can probably ``destroy'' it by the longitudinal ``painting''.

\subsection{Circular attractor in presence of noise}

Now, we will add to our analysis the considerations of the diffusive term in Eq. (\ref{eq2}).

First, let's make some general qualitative observations. 

When $z_{off}=0$, the coherent force $F$ pushes all the protons to the zero amplitude of oscillations, that is, to the center of the bunch. That, according to (\ref{eqNs}), (\ref{eqTau}) and (\ref{eq3b}), increases the rms value of the diffusive kick since the p-bunch density at $\tau=0$ is increasing. The process will go on until the instantaneous cooling and heating rates balance each other. Under assumption of the bunch keeping the Gaussian distribution one can find the analytic expression for the equilibrium $\sigma_\delta$ by equating $\lambda_C$ and $\lambda_D$ in (\ref{eq3d}).

When $z_{off}>z_1$, the coherent force $F$ pulls the protons to the attractor of radius $J_A \equiv \delta_A^2$ (defined by Eq. (\ref{eq5})). That results in a ``hollow'' longitudinal distribution with the number of particles in the central slice approaching a constant value: 

\begin{equation} \label{eq6}
N_{s1}=\frac{N_p z_0 \sigma_{\delta0}}{\pi \sqrt{J_A} \sigma_{ps0}}
\end{equation}
which gives a constant rms amplitude of the dispersive kick

\begin{equation} \label{eq7}
\sqrt{\langle D_1^2 \rangle}=\frac{V_0}{E_0} \sqrt{\frac{N_p \sigma_{\delta0} \sigma_0}{2 \sqrt{2\pi} \delta_A \sigma_{ps0}}} \approx 936 \cdot \frac{V0}{E0}
\end{equation}
Here, the estimate is given for $z_{off}=2$ $\mu$m.

This must be compared to the initial $\sqrt{\langle D_0^2 \rangle}$:

\begin{equation} \label{eq8}
\sqrt{\langle D_0^2 \rangle}=\frac{V_0}{E_0} \sqrt{\frac{N_p \sigma_0}{4 \sigma_{ps0}}} \approx 929 \cdot \frac{V0}{E0}
\end{equation}

\begin{figure}[!htb]
  \centering
  \includegraphics[width=1\columnwidth]{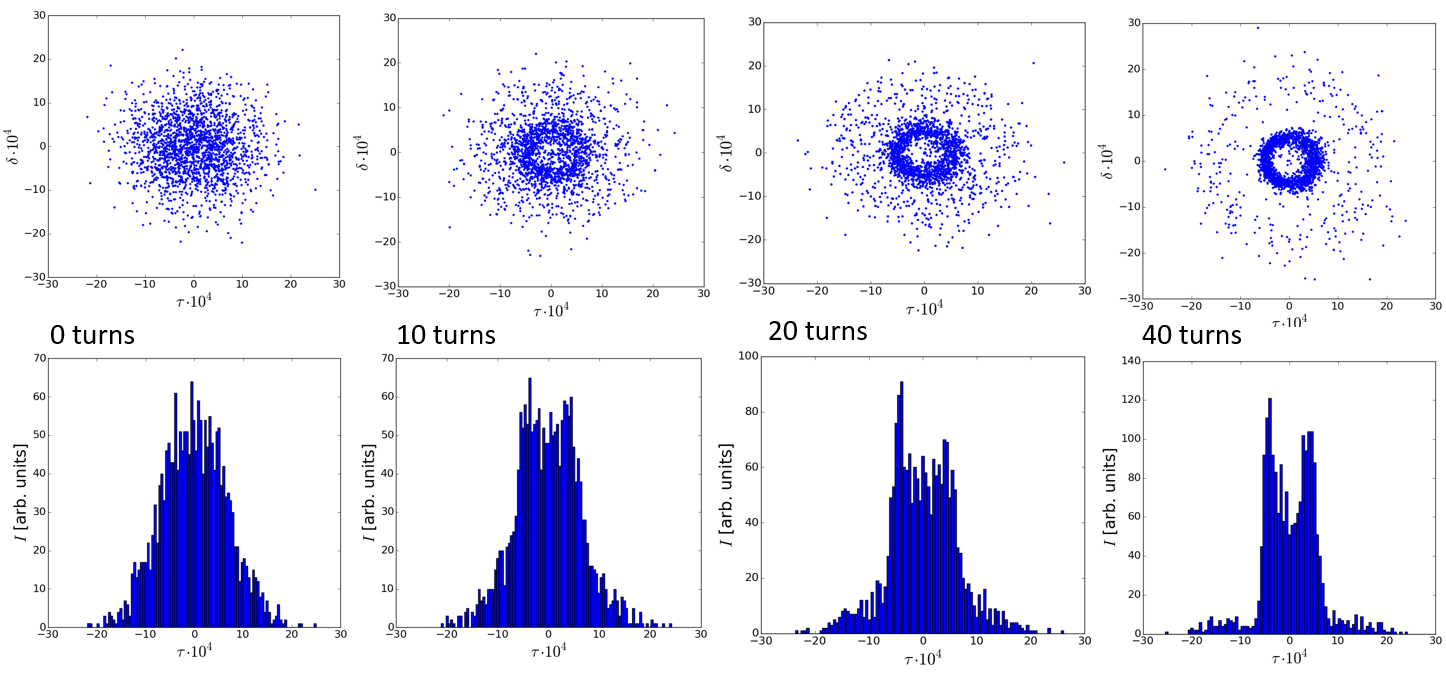}
  \caption{Phase space evolution in the presence of the diffusive kicks with the constant rms value $2\cdot \frac{V_0}{E_0}$.}
   \label{noise}
\end{figure}

Therefore, not only the diffusive kicks don't ``destroy'' the attractor, quite the opposite, the existence of the attractor keeps the noise term constant over time. This means that in the absence of other noise sources a clearly visible doughnut distribution in the phase space will be developed. Just as a simple example, Fig. \ref{noise} shows the evolution of the distribution of the 2000 particles' bunch in the presence of the random kicks with the rms value 2 times larger than the peak coherent force.

In real life, the IBS, a much stronger source of diffusion, will be present. Yet, it won't change the fundamental fact that the presence of the circular attractor will stop the cooling of all the protons with $J<J_A$.

\section{Conclusion}
We found that the circular attractor, which  essentially stops the cooling, will form in the protons' longitudinal phase space if the systematic longitudinal misalignment ($z_{off}$) between the proton and the electron bunches in the kicker section is larger than 1.33 um.

We found the expected ``size'' of the attractor depending on $z_{off}$ (the dependence is given in Fig. \ref{deltaZTrend}).

We discovered that even in the absense of the offset there is a ``weak'' attractor in the longitudinal phase space of the proton bunch, which results from a ``diffusive'' part of the wake and the fact that the cooling e-bunch is much shorter than the p-bunch. Whether the existence of a weak attractor creates a problem for us and how this problem can be solved will be explored in the future studies.   

\section{Acknowledgments}

We wish to thank Michael Blaskiewicz, Sergei Nagaitsev and Gennady Stupakov for the helpful discussions. 

This work was supported by Brookhaven Science Associates, LLC under Contract No. DE-SC0012704 with the U.S. Department of Energy.

\end{document}